\documentclass[doublecol]{epl2} 

\usepackage{graphicx}
\usepackage{float}
\usepackage{amsmath, amssymb}
\usepackage{lmodern}
\usepackage[T1]{fontenc}
\usepackage[utf8]{inputenc}
\usepackage[english]{babel}
\usepackage{color}

\title{Weak species in rock-paper-scissors models}
\shorttitle{Weak species in rock-paper-scissors models} 

\author{P.P. Avelino\inst{1,2}, B.F. de Oliveira\inst{3}, and R.S. Trintin\inst{3}}
\shortauthor{Avelino, de Oliveira, Trintin}

\institute{\inst{1}Departamento de F\'{\i}sica e Astronomia, Faculdade de Ci\^encias, Universidade do Porto, Rua do Campo Alegre s/n, 4169-007 Porto, Portugal\\
\inst{2}Instituto de Astrof\'{\i}sica e Ci\^encias do Espa{\c c}o, Universidade do Porto, CAUP, Rua das Estrelas, 4150-762 Porto, Portugal\\
\inst{3}Departamento de F\'\i sica, Universidade Estadual de Maring\'a, 87020-900 Maring\'a, PR, Brazil\\
}


\abstract{
In this letter, we investigate the population dynamics in a May-Leonard formulation of the rock-paper-scissors game in which one or two species, which we shall refer to as ``weak'', have a reduced predation or reproduction probability. We show that in a nonspatial model the stationary solution where all three species coexist is always unstable, while in a spatial stochastic model coexistence is possible for a wide parameter space. We find, that a reduced predation probability results in a significantly higher abundance of ``weak'' species, in models with either one or two ``weak'' species, as long as the simulation lattices are sufficiently large for coexistence to prevail. On the other hand, we show that a reduced reproduction probability has a smaller impact on the abundance of ``weak'' species, generally leading to a slight decrease of its population size --- the increase of the population size of one of the ``weak'' species being more than compensated by the reduction of the other, in the two species case. We further show that the species abundances in models where both predation and reproduction probabilities are simultaneously reduced may be accurately estimated from the results obtained considering only a reduction of either the predation or the reproduction probability.}

\begin{document}

\maketitle
\section{Introduction}

Ecology is the study of the interactions of living organisms with one another and with their ecosystem. Intra-specific and inter-specific interactions are essential for the coexistence of a wide variety of species and are, therefore, also crucial for development and sustainability of biodiversity. Evolutionary games usually seek to quantify this natural process through the simulation of the dynamics of populations with different species subject to non-hierarchical interactions.

The classic RPS model (see \cite{1920-Lotka-PNAS-6-410, 1926-Volterra-N-118-558, 1975-May-SIAM-29-243} for the pioneer work by Lotka and Volterra, and May and Leonard) describes the dynamics of a population of three species with cyclic predation, as well as mobility and reproduction interactions. In spatial RPS models the coexistence of all three species can be maintained in a significant parameter volume of predation, mobility and reproduction probabilities \cite{2006-Reichenbach-PRE-74-051907, 2007-Perc-PRE-75-052102, 2007-Reichenbach-N-488-1046, 2007-Reichenbach-PRL-99-238105}. Corresponding biological examples have been found in nature, including {\it E. coli} bacteria \cite{2002-Kerr-N-418-171, 2004-Kirkup-Nature-428-412} and lizards \cite{1996-Sinervo-Nature-380-240}.

Population dynamics in spatial stochastic RPS type models has been investigated thoroughly in the literature considering a wide variety of models and perspectives \cite{2008-Szabo-PRE-77-041919, 2011-Allesina-PNAS-108-5638, 2012-Avelino-PRE-86-031119, 2012-Avelino-PRE-86-036112, 2012-Li-PA-391-125, 2012-Roman-JSMTE-2012-p07014, 2013-Lutz-JTB-317-286, 2013-Roman-PRE-87-032148, 2014-Cheng-SR-4-7486, 2014-Szolnoki-JRSI-11-0735, 2016-Kang-Entropy-18-284, 2016-Roman-JTB-403-10, 2017-Brown-PRE-96-012147, 2017-Park-SR-7-7465, 2017-Bazeia-EPL-119-58003, 2017-Souza-Filho-PRE-95-062411, 2018-Shadisadt-PRE-98-062105, 2019-Avelino-PRE-99-052310, 2020-Szolnoki-EPL-131-68001}. Although such dynamics is usually strongly dependent on a variety of model parameters ---  including predation, reproduction and mobility probabilities --- in most cases these are assumed to be the same for all species. However, this simplifying assumption is not expected to constitute a valid approximation in the modelling of most biological systems in nature.

The impact of the reduction of the predation probability of a single (``weakest'') species has been investigated in the context of the spatial stochastic three species RPS model \cite{2001-Frean-PRSLB-268-1323,2009-Berr-PRL-102-048102, 2019-Avelino-PRE-100-042209, 2019-Menezes-EPL-126-18003}. It has been shown that coexistence of all three species prevails for sufficiently large simulation lattices, with the ``weakest'' species being the most abundant \cite{2019-Avelino-PRE-100-042209} --- for smaller simulation lattices, large amplitude oscillations at the initial stages of simulations with random initial conditions can lead to a significant dependence of the chance of ``weakest'' species survival on the lattice size \cite{2019-Avelino-PRE-100-042209,2019-Menezes-EPL-126-18003}. The crucial importance of the simulation lattice size has also been recognized in the context of public goods games with punishment \cite{P1, P2, P3, P4}. Recent research \cite{2020-Liao-N-11-6055} dealing with the inhibition of protein production, the digestion of genomic DNA and the rupture of cellular membrane with three strains of \textit{E. coli} interacting cyclically in an asymmetric system also concluded for the dominance of the ``weakest'' strain.

In this letter we extend previous work on the performance of ``weak'' species, by investigating the dynamics of RPS models in which one or two species have their predation and/or reproduction probability reduced. We start by describing a nonspatial RPS model, showing that it does not allow for stable stationary solutions where all three species coexist. We then describe our May-Leonard implementation of the spatial stochastic RPS model and assess the performance of ``weak'' and ``strong'' species in numerical lattice simulations allowing for the coexistence of all three species. In particular, we shall highlight the dependence of the results on whether the ``weak'' species are characterized by a reduced predation or reproduction probability. We will also describe an approximation allowing for the estimate of the species abundances, in models where both predation and reproduction probabilities are simultaneously reduced, from the results obtained considering only a reduction of either the predation or the reproduction probability. Finally, we present the main conclusions of our work.

\section{Nonspatial RPS model}

Here we shall consider a four-state cyclic predator-prey model, also known as a May-Leonard implementation of the RPS model. The densities of the species $i=1,..,3$ shall be denoted by $\rho_i$. For simplicity of notation, the density associated to the additional state --- denoted by a `0' --- shall be referred to as the density of empty sites, even when considering a nonspatial model. The total density is normalized to unity, so that
\begin{equation}
    \rho_0 + \sum_{i=1}^3 \rho_i = 1\,.
    \label{eq1}
\end{equation}
In a nonspatial RPS model there are two possible interactions: predation [$i\ (i+1) \to i\ 0$] and reproduction [$i\ 0 \to i\ i$], with probabilities $p_i$ and $r_i$, respectively (we shall use modular arithmetic throughout the paper, so that the integers $i$ and $j$ represent the same species whenever $i = j\ {\rm mod} \ 3$, where mod denotes the modulo operation). With an appropriate choice of time unit, the set of equations that represent the evolution of the densities of the different species may be written as 
\begin{eqnarray}
{\dot \rho_i}&=&r_i \, \rho_i \, \rho_{0} - p_{i-1} \, \rho_{i-1} \, \rho_i\,, \label{eq2}
\end{eqnarray}
The stationary solutions to Eq. \eqref{eq2} are
\begin{eqnarray}
\rho_i &=& \dfrac{1}{\left[ 1 + \dfrac{p_i}{r_{i+1}} + \dfrac{p_i r_{i-1}}{p_{i+1}r_{i+1}} + \dfrac{p_i r_i}{p_{i-1}r_{i+1}}\right]}\,, \label{eq4} \\
\rho_0 &=& \dfrac{1}{\left[ 1 + \dfrac{r_1}{p_{3}} + \dfrac{r_{2}}{p_1} + \dfrac{r_{3}}{p_{2}}\right]}\,. \label{eq5}
\end{eqnarray}

We have verified that there are no stationary solutions to Eq.  \eqref{eq2} with $\rho_i \neq 0$ for all $i=1,2,3$ in which the real part of all the eigenvalues of the Jacobian matrix, defined by
\begin{equation*}
\begin{bmatrix}
\frac{\partial \dot{\rho _{1}}}{\partial \rho _{1}} & \frac{\partial \dot{\rho _{1}}}{\partial \rho _{2}} & \frac{\partial \dot{\rho _{1}}}{\partial \rho _{3}}\\
\frac{\partial \dot{\rho _{2}}}{\partial \rho _{1}} & \frac{\partial \dot{\rho _{2}}}{\partial \rho _{2}} & \frac{\partial \dot{\rho _{2}}}{\partial \rho _{3}}\\
\frac{\partial \dot{\rho _{3}}}{\partial \rho _{1}} & \frac{\partial \dot{\rho _{3}}}{\partial \rho _{2}} & \frac{\partial \dot{\rho _{3}}}{\partial \rho _{3}}
\end{bmatrix} =
\end{equation*}
{\footnotesize
\begin{equation*}
\begin{bmatrix}
r_{1}(\rho_{0}-\rho_{1})-p_{3}\rho_{3} & -r_{1}\rho_{1} & -r_{1}\rho_{1}-p_{3}\rho_{1} \\
-r_{1}\rho_{1}-p_{3}\rho_{1} & r_{2}(\rho_{0}-\rho_{2})-p_{1}\rho_{1} & -r_{2}\rho_{2} \\
-r_{3}\rho_{3} & -r_{3}\rho_{3}-p_{2}\rho_{3} & r_{3}(\rho_{0}-\rho_{3})-p_{2}\rho_{2} 
\end{bmatrix}\,,
\end{equation*}}
is negative. Therefore, we conclude that the stationary solutions to Eq. \eqref{eq2}, given by Eqs. \eqref{eq4} and \eqref{eq5}, leading to the coexistence of all three species are always unstable in a nonspatial RPS model\cite{Hofbauer1998}.

\section{May-Leonard spatial stochastic RPS model}

In this letter we shall consider a May-Leonard implementation of the spatial RPS model on a two-dimensional square lattice with $\mathcal{N}$ sites and periodic boundary conditions. At each simulation time step, an individual (active) is randomly chosen to interact with one of its four nearest neighbors (passive) --- also randomly selected. Then, one of three possible interactions (predation, reproduction, or mobility) is randomly chosen (with probabilities $p$, $m$ and $r$, respectively). In this paper we shall consider the so-called von Neumann neighbourhood (or 4-neighbourhood) composed of a central cell (the active one) and its four non-diagonal adjacent cells. Whenever mobility is selected the active swaps place with the passive. On the other hand, if reproduction is selected and the passive is an empty site (otherwise the interaction cannot be completed, and the procedure is repeated once again), then the site becomes occupied with an individual of the same species of the active. Finally, if predation is selected and the passive is a prey of the active (otherwise the interaction cannot be completed, and the procedure is repeated once again), then the passive is removed from the lattice and an empty site is left at its previous position. A time step is completed whenever one interaction is carried out. A generation time (our time unit) corresponds to $\mathcal{N}$ time steps.

Here we shall investigate the dynamical impact of a reduction of the predation or reproduction probabilities of one or two species by a factor of $\mathcal{P}_{w} \in \ ]0,1]$ or $\mathcal{R}_{w} \in \ ]0,1]$, respectively. We will consider the following cases: Model $P_1$ - one species with reduced predation probability, $p_1 = \mathcal{P}_w p$ ($p_2= p_3 = p$); Model $R_1$ - one species with reduced reproduction probability $r_1 = \mathcal{R}_w r$ ($r_2= r_3 = r$); Model $P_2$ two species with reduced predation probabilities $p_1 = p_2 = \mathcal{P}_w p$ ($p_3 = p$); Model $R_2$ - two species with reduced reproduction probabilities $r_1 = r_2 = \mathcal{R}_w r$ ($r_3 = r$). In the models $P_1$ and $P_2$ we take $r_1 = r_2 = r_3 = r$. In the models $R_1$ and $R_2$ it is assumed that $p_1 = p_2 = p_3 = p$. We shall also consider two additional models, $(PR)_1$ and $(PR)_2$ (with one and two ``weak'' species, respectively), in which both the predation and reproduction probabilities are reduced. 

\begin{figure}[!ht]
	\centering
	\includegraphics[scale=0.75]{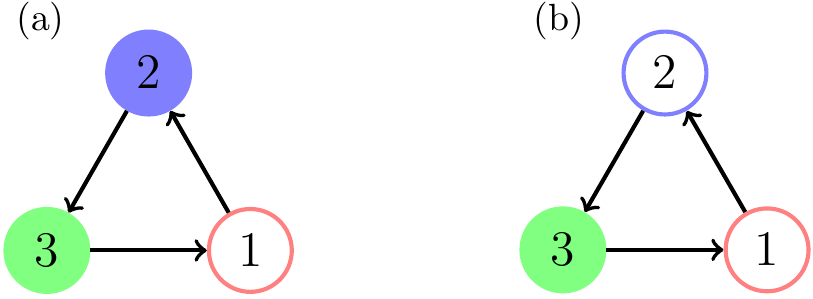}
	\caption{Scheme of predation interactions of the $3$ species RPS model: (a) represents the case with a single ``weak'' species (models $P_1$, $R_1$, $(PR)_1$); (b) represents the case with two `weak'' species (models $P_2$, $R_2$, $(PR)_2$). Filled and unfilled circles represent ``weak'' and ``strong'' species, respectively.}
	\label{fig1}
\end{figure}

Fig.~\ref{fig1} displays the scheme of cyclic predator-prey interactions of the $3$ species RPS models we investigate in this letter, with the filled and unfilled circles representing ``weak'' and ``strong'' species, respectively --- cases (a) and (b) correspond to models with one and two ``weak'' species. As previously mentioned, the six models considered in this letter (three with one ``weak'' species and three with two ``weak'' species) are denoted by $P_n$, $R_n$ and $(PR)_n$, depending, respectively, on whether only the predation rate is reduced, only the reproduction rate is reduced or both --- $n$ represents the number of ``weak'' species ($n=1$ or $n=2$). Model $P_1$ has also been investigated in \cite{2019-Avelino-PRE-100-042209}.

\section{Results}
Here we assess the performance of ``weak'' and ``strong'' species using spatial stochastic simulations of the RPS models $P_n$, $R_n$ and $(PR)_n$, with $n=1,2$. 

\begin{figure}[!ht]
	\centering
	\includegraphics{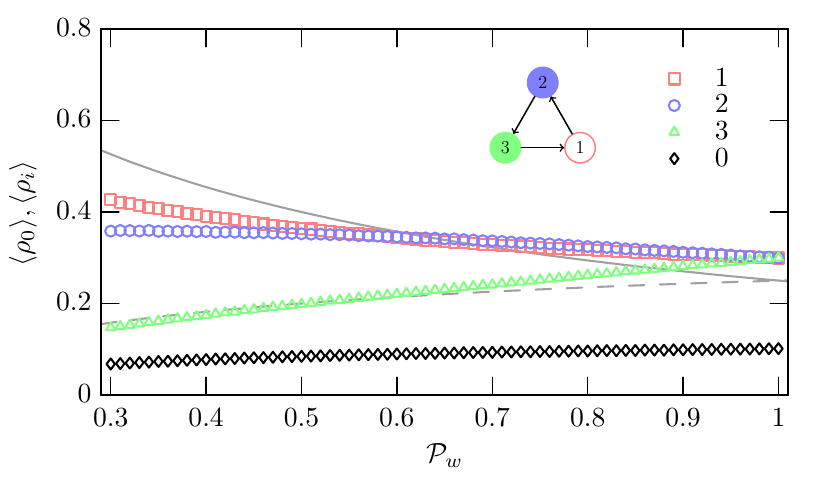}
	\caption{The average density of the species ($i=1,2,3$) and empty sites ($i=0$) for the spatial model $P_1$, as a function of $\mathcal{P}_w$, assuming $m = 0.5$ and $p = r = 0.25$. Notice that the ``weak'' species and its prey are the most abundant. The grey lines represent the stationary nonspatial solution, given in Eqs. \eqref{eq4} and \eqref{eq5}, for the density of the species 1 (solid line), 2 and 3 and empty sites (dashed line). Notice the  significant differences between spatial and nonspatial results (this is also be verified in Figs.~\ref{fig3}, ~\ref{fig4}, and ~\ref{fig5}).}
	\label{fig2}
\end{figure}

Fig.~\ref{fig2} shows the value of the average density of the three species and empty sites as a function of $\mathcal{P}_w$ for the model $P_1$ with a single ``weak'' species. $\mathcal{P}_w$, represents the reduction of the predation probability of the ``weak'' species with respect to the standard case in which the predation probability is the same for all species. The data points result from an average over the last $10^4$ generations of a simulation with a time span equal to $1.5\times 10^4$ generations performed on a $1000^2$ lattice, large enough to guarantee the preservation of coexistence. The results for $\mathcal{P}_w = 1$ were computed starting from random initial conditions. To ensure fast convergence we used the final conditions of the simulations with $\mathcal{P}_w = 1$ as initial conditions of new simulations with $\mathcal{P}_w = 1 - 0.01$. The same procedure was repeated until $\mathcal{P}_w = 0.3$ was reached. We verified that, with such conditions, $5\times 10^3$ generations are sufficient for $\langle \rho_i \rangle$ and $\langle \rho_0 \rangle$ to attain their asymptotic value. Fig.~\ref{fig2} shows that the ``weak'' species generally has an advantage, specially over its predator (the ``weak'' species abundance is similar to that of its prey for $\mathcal{P}_w \in [0.5,1]$). The lines represent the nonspatial stationary analytical (unstable) solution, given in Eqs. \eqref{eq4} and \eqref{eq5}, for the density of the species 1 (solid line), and 2 and 3 and empty sites (dashed line).

\begin{figure}[!ht]
	\centering
	\includegraphics{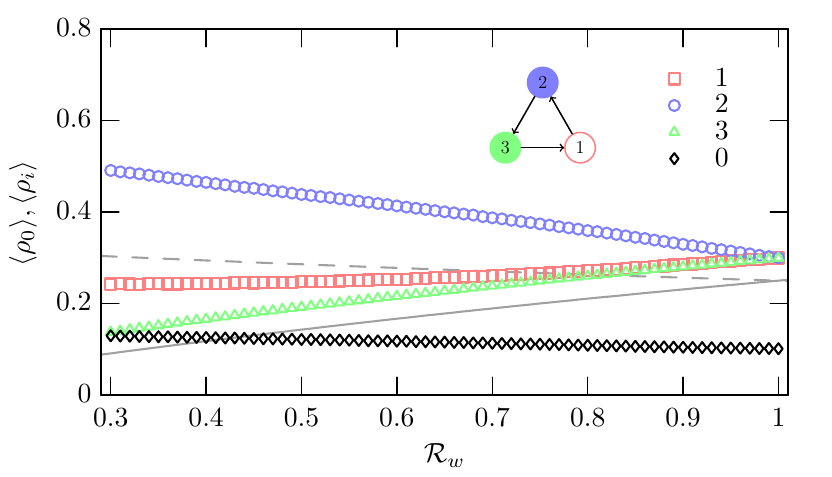}
    \caption{The average density of the species ($i=1,2,3$) and empty sites ($i=0$) for the spatial model $R_1$, as a function of ${\mathcal R}_w$, assuming $m = 0.5$ and $p = r = 0.25$. In this case the performance of the  ``strong'' species surpasses, on average, that of the ``weak'' species. The grey lines represent the stationary nonspatial solution, given in Eqs. \eqref{eq4} and \eqref{eq5}, for the density of the species 1 (solid line), 2 and 3 and empty sites (dashed line).}
    \label{fig3}
\end{figure}

Fig.~\ref{fig3} is analogous to Fig.~\ref{fig2}, except that in Fig.~\ref{fig3} model $R_1$ is considered, with $\mathcal{R}_w$ representing the reduction of the predation probability of the ``weak'' species with respect to the standard case in which the predation probability is the same for all species. Fig.~\ref{fig3} shows that in this case the density of the ``weak'' species is always smaller than the average density of the ``strong'' species. The relatively small advantage that the ``weak'' species has over its predator does not compensate the disadvantage with respect to its prey.

\begin{figure}[!ht]
	\centering
	\includegraphics{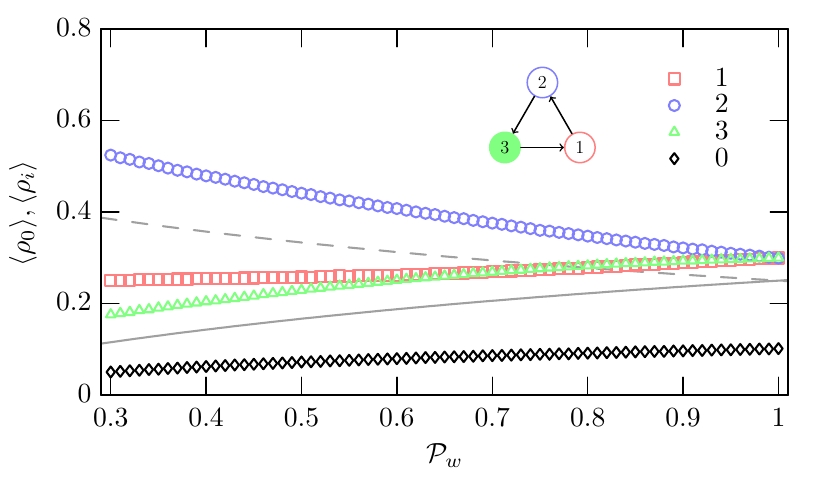}
	\caption{The same as in Fig.~\ref{fig2}, but now for the spatial model $P_2$ with two ``weak'' species having a lower predation probability. Notice that both ``weak'' species are more abundant than the ``strong'' species for $\mathcal{P}_{w} \lesssim 0.6.$. The grey lines represent the stationary nonspatial solution, given in Eqs. \eqref{eq4} and \eqref{eq5}, for the density of the species 1 and 2 (dashed line), and 3 and empty sites (solid line). }
	\label{fig4}
\end{figure}

Fig.~\ref{fig4} displays the results of the model $P_2$. In this model there are two ``weak'' species which have a lower predation probability. Fig.~\ref{fig4} shows  that, in this model, the two ``weak'' species ($1$ and $2$, but specially species $2$) generally have advantage over the ``strong'' species ($3$). This shows that the advantage of the ``weak'' species found in model $P_1$ is also observed in three species RPS models with two ``weak'' species.

\begin{figure}[!ht]
	\centering
	\includegraphics{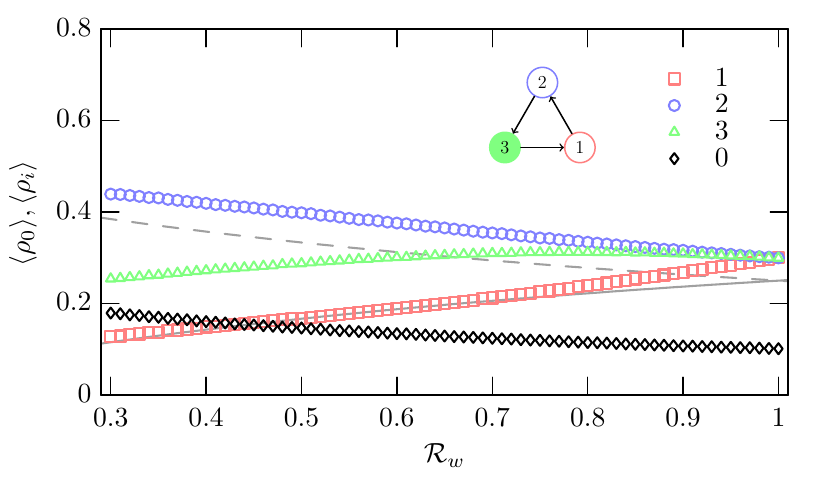}
	\caption{The same as in Fig.~\ref{fig3}, but now for the spatial model $R_2$. In this case one weak species is the most abundant and the other is the least  abundant. The grey lines represent the stationary nonspatial solution, given in Eqs. \eqref{eq4} and \eqref{eq5}, for the density of the species 1 and 3 (solid line), and 2 and empty sites (dashed line).}
	\label{fig5}
\end{figure}

Fig.~\ref{fig5} displays the results of the model $R_2$. This model has two ``weak'' species characterized by a lower reproduction rate. Fig.~\ref{fig5} shows that, in this model, only one of the two ``weak'' species ($2$, the predator of the ``strong'' species) has an advantage over the ``strong'' species ($3$). However, contrary to the case of models $P_1$ and $P_2$, and in agreement with model $R_1$, for ${\mathcal R}_w > 0.5$ there is on average no advantage in being the ``weak'' species.

The discrepancies between spatial and nonspatial solutions shown in Figs.~\ref{fig2}, \ref{fig3}, \ref{fig4}, \ref{fig5}  for models $P_1$, $R_1$, $P_2$ and $R_2$, respectively, are significant. They are a demonstration of the importance of the spatial structure, such as the spiral patterns that emerge on this type of models, on the dynamics of population networks. It is this spatial structure which is responsible for the stability of the population of three species in sufficiently large spatial simulations, in sharp contrast with the instability associated to the nonspatial stationary solutions.

\begin{figure}[!ht]
	\centering
	\includegraphics{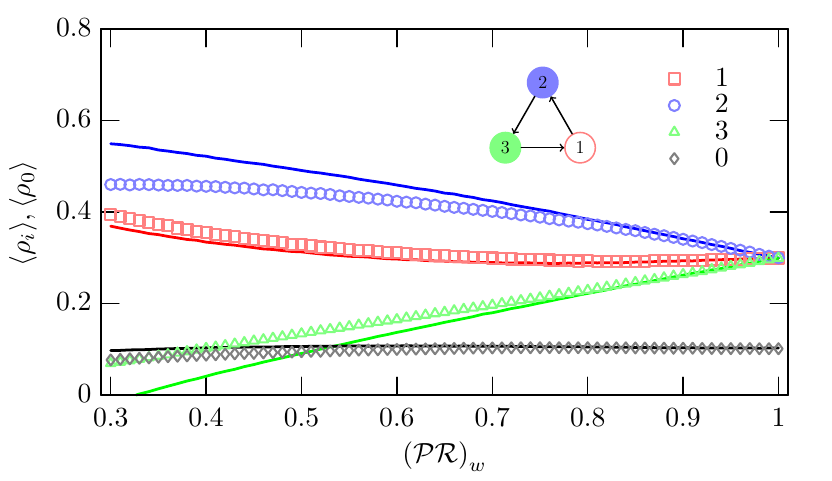}
	\caption{The average density of the species ($i=1,2,3$) and empty sites ($i=0$) for the model $(PR)_1$, as a function of $(\mathcal{PR})_w = \mathcal{P}_w = \mathcal{R}_w$, assuming $m = 0.5$ and $p = r = 0.25$ (in this model both the predator and reproduction probabilities are reduced by the same amount). The lines representing the approximate solution obtained using Eqs. \eqref{approx_i} and \eqref{approx_0}, in combination with the results of models $P_1$ and $R_1$, provide an excellent fit to the results of model $(PR)_1$.}
	\label{fig6}
\end{figure}

Fig.~\ref{fig6} displays the results of the model $(PR)_1$ with one ``weak'' species ($1$). These results were obtained using the same procedure followed in Figs.~\ref{fig2} and Figs.~\ref{fig3} for models $P_1$ and $R_1$, respectively, but now reducing both the predation and reproduction of the ``weak'' species ($1$) by the same amount  ($(\mathcal{PR})_w ={\mathcal P}_w = {\mathcal R}_w$). Fig.~\ref{fig6} shows that, in this case, one of the strong species the species ($2$) is the most abundant followed by the ``weak'' species ($1$). The ``weak'' species is therefore not the most abundant, but its population is always larger than that of its predator.

Motivated by the fact that, for fixed $p$, $r$ and $m$, and $1-{\mathcal P}_w \ll 1$ and $1-{\mathcal R}_w \ll 1$
\begin{eqnarray}
\rho ({\mathcal P}_w,{\mathcal R}_w) &\sim& \rho (1,1) + \left.\frac{\partial \rho}{\partial {\mathcal P}_w}\right|_{(1,1)} (\mathcal{P}_w-1) \nonumber \\ &&+\left.\frac{\partial \rho}{\partial {\mathcal R}_w}\right|_{(1,1)} (\mathcal{R}_w-1)\nonumber \\
&\sim& \rho ({\mathcal P}_w,1)+\rho (1,{\mathcal R}_w)-\rho (1,1)\,.
\end{eqnarray}
We consider the following approximation, allowing for the estimation of the abundances of the species and empty sites in models where both predation and reproduction probabilities are simultaneously reduced from the results obtained considering only a reduction of either the predation or the reproduction probability:

\begin{equation}
\rho_i^{\rm approx} ({\mathcal P}_w,{\mathcal R}_w) =  \rho_i ({\mathcal P}_w,1)+\rho_i (1,{\mathcal R}_w)
-\rho_i (1,1)\,,
\label{approx_i}
\end{equation}

\begin{equation}
\rho_0^{\rm approx} ({\mathcal P}_w,{\mathcal R}_w) =  \rho_0 ({\mathcal P}_w,1)+\rho_0 (1,{\mathcal R}_w)
-\rho_0 (1,1)\,.
\label{approx_0}
\end{equation}

The lines in Fig.~\ref{fig6} show the results obtained using the approximation given in Eqs. \eqref{approx_i} and \eqref{approx_0} (i.e., summing the results of models $P_1$ and $R_1$ for the corresponding values of $(\mathcal{P})_w$ and $(\mathcal{R})_w$ and then subtracting either $\langle \rho_i((\mathcal{PR})_w = 1) \rangle = 0.3$ [species $1$, $2$ and $3$] or $\langle \rho_0((\mathcal{PR})_w = 1) \rangle = 0.1$ [empty sites]). The quality of the approximation is remarkable, especially for $(\mathcal{PR})_w > 0.6$, thus showing that an accurate estimate of the model $(PR)_1$ results may be obtained directly from those found for models $P_1$ and $R_1$. The more significant differences observed for $(\mathcal{PR})_w < 0.6$ are a further demonstration of the non-linear behaviour of these systems.

\begin{figure}[!ht]
	\centering
	\includegraphics{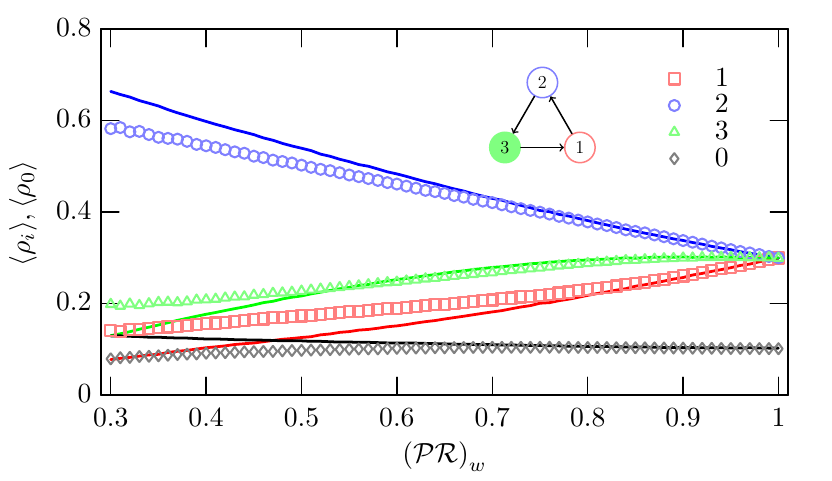}
	\caption{The same as in Fig.~\ref{fig6}, but now for the model $(PR)_2$. These results show that the approximate solution, obtained using Eqs. \eqref{approx_i} and \eqref{approx_0} in combination with the results of models $P_2$ and $R_2$, provides an excellent fit to the results of model $(PR)_2$.}
	\label{fig7}
\end{figure}

Fig.~\ref{fig7} is similar to Fig.~\ref{fig6}, but now for the model $(PR)_2$ (in this model the predation and reproduction probabilities are reduced by the same amount for species $1$ and $2$). Fig.~\ref{fig7} shows that, in this case, one of the ``weak'' species ($2$) is the most abundant followed by the ``strong'' species ($3$), with the remaining ``weak'' species ($1$) being the least abundant. Again, it is possible to verify that the approximate solution, obtained using Eqs. \eqref{approx_i} and \eqref{approx_0} in combination with the results of models $P_2$ and $R_2$, provides an excellent fit to the results of model $(PR)_2$. We have also verified that the accuracy of this approximation is, in general, much better than $20 \%$ for models with reduced predation and reproduction probabilities (not necessarily by the same amount) in the range $1 \le {\mathcal P}_w+{\mathcal R}_w \le 2$.

\section{Conclusions}

In this letter we investigated the performance of ``weak'' and ``strong'' species in a May-Leonard formulation of a three species cyclic predator-prey model. We generalized previous work by considering cases with two ``weak'' species and by extending the definition of ``weak'' species to also include species with a reduced reproduction rate. We have shown that the spatial dynamics is essential for coexistence, with stationary solutions in which all three species coexist being always unstable in nonspatial models. On the other hand, we have found that, for sufficiently large simulation lattices, all three species may naturally coexist for a wide range of parameters. We have shown that the significantly higher abundance of ``weak'' species found in spatial stochastic RPS models with a single ``weak'' species also occurs in models with two ``weak'' species. However, we have shown that the predominance of the ``weak'' species does not generally hold if the ``weak'' species have a reduced reproduction rate. We have also developed a simple approximation which allows for an  accurate estimation of the species abundances in models where both predation and reproduction probabilities of ``weak'' are simultaneously reduced from the results obtained considering only a reduction of either the predation or the reproduction probability.

\acknowledgments
P.P.A. acknowledges the support by Fundação para a Ciência e a Tecnologia (FCT) through the research grants UIDB/04434/2020, UIDP/04434/2020. B.F.O. and R.S.T. thanks CAPES-Finance Code 001, Funda\c c\~ao Arauc\'aria, and INCT-FCx (CNPq/FAPESP) for financial and computational support.


\end{document}